\documentclass[notitlepage,floatfix,aps,prapplied,reprint,amsmath,amssymb,twocolumn,superscriptaddress,10pt,floatfix]{revtex4-1}
\usepackage{graphicx}
\usepackage{dcolumn}
\usepackage{bm}
\usepackage{braket}
\usepackage{amssymb}
\usepackage{hyperref}
\usepackage{amsmath}
\usepackage{textcomp}
\usepackage{amsfonts}
\usepackage{gensymb}

\begin{document}

\title{Multiplexed Single Photons from Deterministically Positioned Nanowire Quantum Dots}
\author{Zhe-Xian Koong}
\email[Correspondence: ]{zk49@hw.ac.uk}
\affiliation{
 SUPA, Institute of Photonics and Quantum Sciences, Heriot-Watt University, EH14 4AS, United Kingdom
}
\author{Guillem Ballesteros-Garcia}
\author{Raphaël Proux}
\affiliation{
 SUPA, Institute of Photonics and Quantum Sciences, Heriot-Watt University, EH14 4AS, United Kingdom
}
\author{Dan Dalacu}
\affiliation{National Research Council of Canada, Ottawa, Ontario, K1A 0R6, Canada}
\author{Philip J. Poole}
\affiliation{National Research Council of Canada, Ottawa, Ontario, K1A 0R6, Canada}
\author{Brian D. Gerardot}
\email[Correspondence: ]{b.d.gerardot@hw.ac.uk}
\affiliation{
 SUPA, Institute of Photonics and Quantum Sciences, Heriot-Watt University, EH14 4AS, United Kingdom
}

\date{\today}

\begin{abstract}
Solid-state quantum emitters are excellent sources of on-demand indistinguishable or entangled photons and can host long-lived spin memories, crucial resources for photonic quantum information applications. However, their scalability remains an outstanding challenge. 
Here we present a scalable technique to multiplex streams of photons from multiple independent quantum dots, on-chip, into a fiber network for use "off-chip”. 
Multiplexing is achieved by incorporating a multi-core fiber into a confocal microscope and spatially matching the multiple foci, seven in this case, to quantum dots in an array of deterministically positioned nanowires. 
First, we report the coherent control of the emission of biexciton-exciton cascade from a single nanowire quantum dot under resonant two-photon excitation. 
Then, as a proof-of-principle demonstration, we perform parallel spectroscopy on the nanowire array to identify two nearly identical quantum dots at different positions which are subsequently tuned into resonance with an external magnetic field. 
Multiplexing of background-free single photons from these two quantum dots is then achieved. 
Our approach, applicable to all types of quantum emitters, can readily be scaled up to multiplex $>100$ quantum light sources, providing a breakthrough in hardware for photonic based quantum technologies. 
Immediate applications include quantum communication, quantum simulation, and quantum computation. 
\end{abstract}

\maketitle

\section{Introduction}

Motivated by the phenomenal scalability of semiconductor integrated circuits for classical computing and communication, semiconductor quantum photonic chips have been pursued for light-based quantum information technologies. 
Visions of fully integrated quantum photonic chips, which require on-demand indistinguishable single photon generation integrated on-chip with low-loss directional couplers, phase shifters, filters, and single photon detectors, have been presented~\cite{lodahl_interfacing_2015,dietrich_gaas_2016,bogdanov_material_2017,kim_hybrid_2020}. 
On-chip integration of solid-state emitters such as color centers in diamond~\cite{mouradian_scalable_2015,chen_laser_2016,machielse_quantum_2019,wan_large-scale_2019}, molecules~\cite{grandi_hybrid_2019,rattenbacher_coherent_2019}, 2D materials~\cite{branny_deterministic_2017,froch_coupling_2019,Peyskens2019,herranz_on_2020}, and III-V semiconductor quantum dots (QDs)~\cite{somaschi_near-optimal_2016,ding_-demand_2016,liu_solid-state_2019,coles_chirality_2016,davanco_heterogeneous_2017,grim_scalable_2019,uppu_scalable_2020}, are particularly promising for these applications. As solid-state emitters reach maturity, the next logical step is to interface these sources with larger quantum photonic architectures to promote scalability and realization of multipartite quantum information protocols~\cite{slussarenko_photonic_2019}.  
For example, near term application of quantum information protocols such as quantum key distribution ~\cite{takemoto_quantum_2015} (including measurement-device-independent  schemes~\cite{lo_measurement_2012}), boson sampling~\cite{loredo_boson_2017,wang_boson_2019} and photonic cluster state generation for measurement-based quantum computing~\cite{schwartz_deterministic_2016,scerri_frequency-encoded_2018,gimeno-segovia_deterministic_2019} all benefit from having multiple streams of indistinguishable photons that can be realized with QDs. 
However, while each independent ingredient of the fully integrated chip vision has been realized separately, hybrid integration of all components in a fully functional platform is a demanding long-term challenge. 
Hence, in the near term, a viable approach to develop applications is to connect the separate components via optical interlinks such as optical fibers. 
In addition, long-term applications such as long-distance quantum communication~\cite{duan_long-distance_2001} or distributed quantum networking~\cite{cirac_distributed_1999,kimble_quantum_2008} necessitate coupling the quantum light off-chip into a fiber network. 

To handle the capacity and connectivity required for these applications, multiplexing techniques which expand  the current solid-state hardware infrastructure are required. One approach is the integration of multi-core fibers (MCFs) in a fiber-based quantum optical network. MCFs have $n$ cores within a single cladding and achieve single-mode transmission of each core with minimal crosstalk~\cite{hayashi_design_2011}. These properties enable ultra-high (Tb/s) optical transmission capacity~\cite{sakaguchi_2011} and spatial-division multiplexing, with up to $n=120$ cores~\cite{birks_photonic_2012} for optical quantum communication~\cite{ding_high-dimensional_2017}. 

In this article, we introduce a novel multi-spot microscopy setup which combines two MCFs for individual excitation and collection of multiple foci in a confocal microscope. By matching the foci from each fiber core to a custom designed chip of spatially positioned nanowire arrays with embedded QDs, we are able to probe multiple independent quantum emitters (up to $n=7$) on a single chip and collect the fluorescence from each quantum emitter into a separate single mode fiber. 
By connecting the MCF to a linear fiber array at the input of an imaging spectrometer, we are able to simultaneously perform spectroscopy on the seven independent QDs in the array. 
We take advantage of this to rapidly scan the quantum emitter array and identify two QDs with ground state emissions separated by just 0.113~meV, which we subsequently tune into resonance with an external magnetic field. 
By exciting both QDs individually, we observe suppressed multi-photon emission probability in the emission from both QDs, which validates the generation of single photons from remote emitters on the same chip. 
This constitutes a proof-of-principle demonstration of multiplexing single photon streams from separate quantum sources. 
Additionally, we report on the coherent control of the population of the biexciton and exciton states under resonant two-photon excitation, on a single nanowire QD. 
As a consequence of coherent driving, we show signatures of coherent state manipulation in the biexciton-exciton cascade system, i.e. the observation of Autler-Townes splitting and Rabi oscillation under continuous wave and pulsed driving, respectively. 
Our results motivate the integration of both coherent quantum state manipulation techniques and the multi-spot microscope to generate streams of indistinguishable single photons from multiple emitters on the same chip, which are crucial resources for photonic quantum technologies. 

\begin{figure}
    \includegraphics[width=0.5\textwidth]{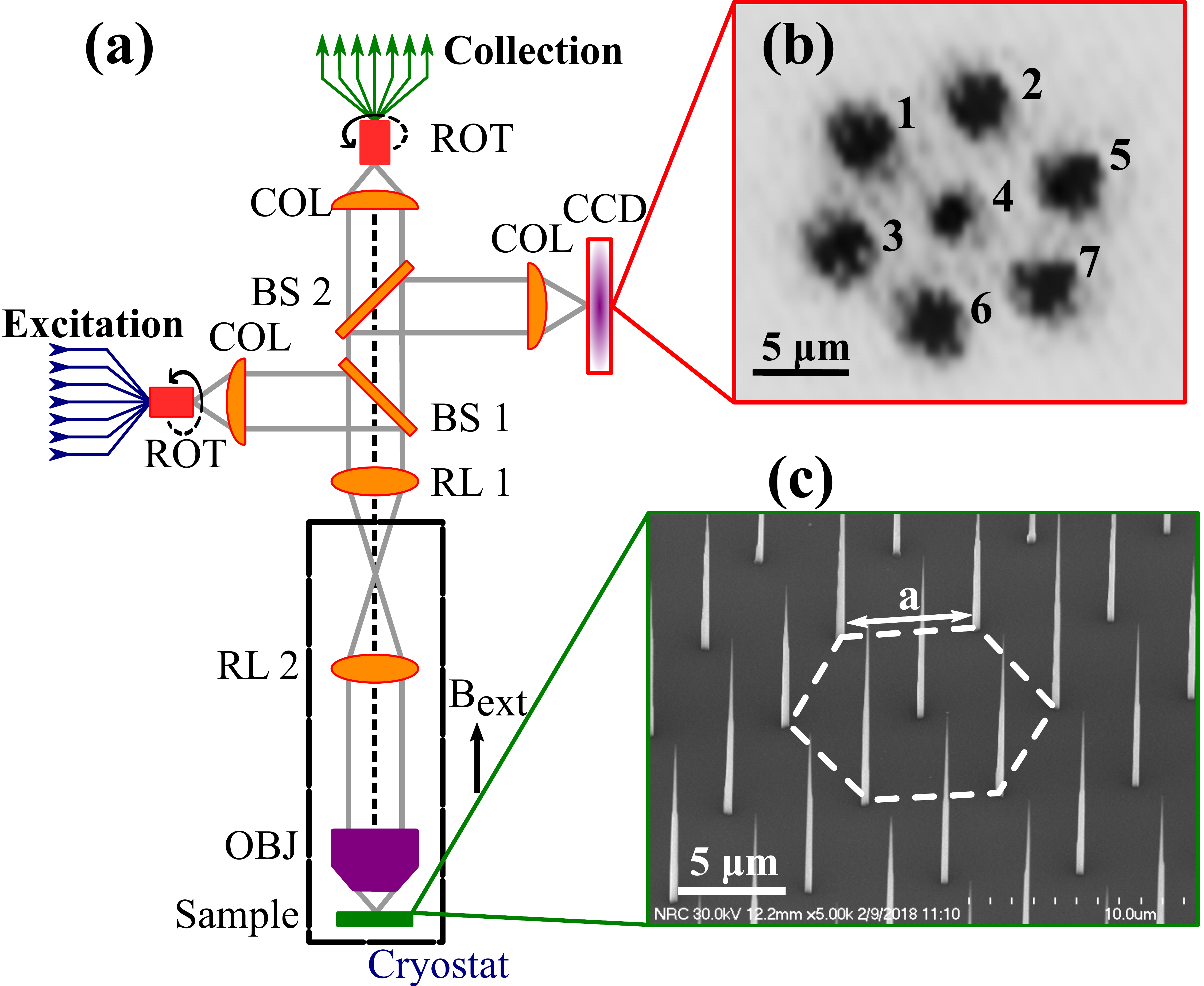}  
    \caption{\textbf{Multi-spot microscope and nanowire sample structure.}
    \textbf{(a)} Schematic of the lens configuration of the multi-spot microscope. COL: collimating lens, ROT: rotatable fiber mount,  BS: 96:4 beam splitter, RL: relay lens, OBJ: microscope objective. CCD: charged-coupled device. 
    \textbf{(b)} Image of the collection/excitation spots on the CCD device.
    \textbf{(c)} SEM image of the tilted sample. The nanowires are arranged in a hexagonal lattice (the geometry is outlined by the dashed lines) with lattice spacing $a$. 
    } 
    \label{fig1} 
\end{figure}

\section{Low-temperature multi-spot confocal microscope setup}
Figure~\ref{fig1}(a) shows the schematic of the multi-spot microscope. We use two commercial MCFs (OptoScribe, COF073400, core diameter of $8.2\,\mu\mathrm{m}$) consisting of \textit{n} = 7 cores arranged in a hexagonal pattern with a lattice constant of $a_0=35\,\mu \mathrm{m}$). 
The fan-outs connected to each of the MCFs, at the excitation and collection, allow us to excite and collect from each focus independently. 
Our microscope builds on a previous design using only a MCF in the collection arm of the microscope~\cite{munoz-matutano_parallel_2016} that prevented independent excitation and collection at each focus. The MCFs are mounted on rotation cages to allow precise alignment of the excitation and collection spots. The lateral displacement of the beam from the MCF, collimated using a collimating lens (Thorlabs C280TMD-B, focal length of $f_{1}=18.4\,\mathrm{mm}$ and NA = 0.15), is compensated with a pair of relay lenses (RL1 and RL2) to form a quasi-4f configuration. The relay lenses, both singlets, have focal lengths of $f_{2}=250\,\mathrm{mm}$ (RL1) and $f_{3}=300\,\mathrm{mm}$ (RL2). They are positioned 50~cm above the microscope objective (focal length of $f_{obj}=2.93\,\mathrm{mm}$ and NA = 0.81). 
This gives an estimated magnification of $M=f_{obj}f_3/f_1f_2\approx 0.19$ and an estimated spot size separation of $a = Ma_0\approx 6.7\,\mu\mathrm{m}$. Upon illumination of all 7 cores of the MCF using the same 940~nm (1.319~eV) narrow-band laser, we observe an image of the MCF on the camera (CCD). The image, shown in Figure~\ref{fig1}(b), gives a measured spot spacing of $6.5\,\mu\mathrm{m}$, with the slight deviation from the estimated value due to a small inaccuracy in the position of the relay lenses.  

In this work, we utilize InP nanowires deterministically positioned in a hexagonal lattice with lattice constant ($a$) ranging from $6.1\,\mu\mathrm{m}$ to $7\,\mu\mathrm{m}$. 
Figure~\ref{fig1}(c) shows a scanning electron microscope (SEM) image of the tilted nanowire sample. The range of lattice constants is designed to optimize the spatial overlap between the foci of the multi-core fiber to the position of the nanowires when accounting for a small deviation in the magnification from the expected value. 
Within each nanowire, InAsP QDs with ground state emission from 910 to 950~nm are embedded.
Further details of the growth process are described in Refs.~\cite{dalacu_selective-area_2009,poole_interplay_2012}.
The sample is kept at $T=3.7\,\mathrm{K}$ in a closed-cycle cryostat (Attocube attoDry1000) equipped with a 9~T superconducting magnet. 
The sample is mounted in the Faraday configuration, where the external magnetic field ($\mathrm{B_{ext}}$) is parallel to the nanowire growth axis. 
For antibunching experiments, a tunable band pass filter with a full-width at half maximum of 1~nm (1.47 meV), centered at 916.30~nm (1.3531~eV), is placed before the collimating lens at the collection to filter out the above-band continuous wave (CW) excitation laser at 830~nm (1.4938~eV) and any unwanted emission lines.
These filtered photons, coupled into the MCF in the collection, are directed to a Hanbury-Brown and Twiss (HBT)-type interferometer.
Here, the fan-out coupler with the collected photon streams are sent to a 50:50 non-polarizing fiber beam splitter.
The fiber beam splitter outputs are then sent to a pair of superconducting nanowire single photon detectors (SNSPD), each with a nominal timing jitter and detection efficiency of $\sim 100\,\mathrm{ps}$ and $\sim 90\%$ at 950~nm, respectively.

\begin{figure*}
    \includegraphics[width=1\textwidth]{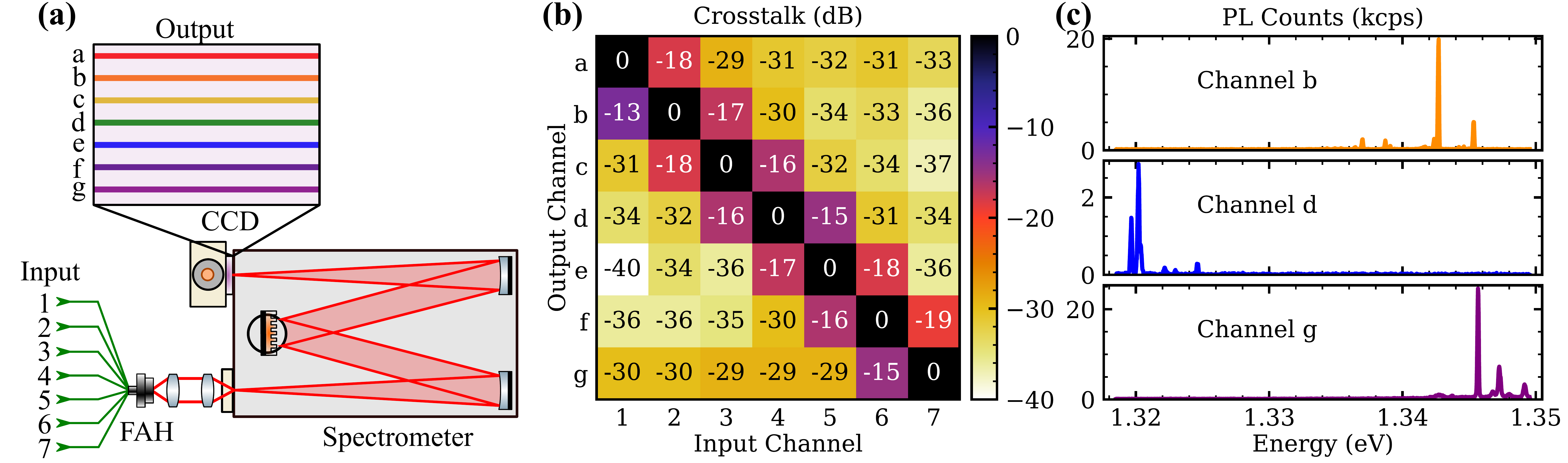}
    \caption{
        \textbf{Multiplexed spectroscopy.}
        \textbf{(a)} Schematic of the setup for fluorescence detection on the spectrometer. 
        The MCF fan-outs (labeled from 1 to 7) are coupled into an imaging spectrometer via a fiber array holder (FAH).
        The multi-spot signals are imaged on a CCD detector array (1340 x 100 pixels), with a separate region of interest, labeled from \textit{a} to \textit{f}, indicated on the CCD.
        \textbf{(b)} Crosstalk matrix between the input fiber fan-out and the output pixels on the spectrometers.
        The adjacent channels show crosstalk of $\approx -15\,\mathrm{dB}$, while the rest of the channels show crosstalk of $\approx -35\,\mathrm{dB}$, approaching the dark counts of the spectrometer.
        \textbf{(c)} An example of the emission spectra from 3 nanowires (corresponding to fluorescence at channel \textit{b}, \textit{e} and \textit{g}) measured simultaneously.  
    }
    \label{fig3a}
\end{figure*}

To analyze the signal from all seven outputs from the collection MCF simultaneously on a spectrometer, we mount the MCF fan-out to a custom-designed fiber array holder (FAH) in front of the entrance slit of an imaging spectrometer. The schematic of the fluorescence detection setup is illustrated in Figure~\ref{fig3a}(a).
The FAH ensures that all of the outputs from the microscope (labeled 1 to 7) can be imaged on the CCD array (1340 x 100 pixels) simultaneously, with their corresponding region of interests on the CCD array (labeled as \textit{a} to \textit{f}) vertically displaced relative to each other.
We characterize the crosstalk matrix of the full system (the multi-spot microscope and the imaging spectrometer setup combined) by injecting laser light into the microscope through one of the excitation fibers and measure the signal received at each of the detection spots on the spectrometer. 
The crosstalk matrix for all seven cores are summarized in Figure~\ref{fig3a}(b). 
Despite showing near dark counts performance (nominal crosstalk of $\approx -35\,\mathrm{dB}$) for most of the pixels on the CCD, the crosstalk measured on adjacent pixels indicates a non-ideal nominal $\approx -15\,\mathrm{dB}$ leakage between adjacent spots.
The crosstalk between channels stems from imperfect focus at the sample that arises from the multi-mode behavior of each fiber core at $\lambda = 940\,\mathrm{nm}$, which inevitably couples scattered laser light into neighboring spots, as well as due to the leakage between adjacent regions of interest on the spectrometer's CCD array. 
Here, we partially mitigate this by illuminating non-adjacent excitation/collection spots where the detection suffers from the least crosstalk. 
By illuminating three out of seven of the excitation spots, we can perform clean spectroscopy on three spatially separated nanowire QDs simultaneously, as shown in Figure~\ref{fig3a}(c). 
This demonstrates the ability to collect fluorescence from multiple nanowire QDs concurrently using the multi-spot microscope and the spectroscopy setup successfully achieves spatial multiplexing from multiple emitters.

\section{Spectroscopy and coherent control of nanowire quantum dots}
\begin{figure*}
    \includegraphics[width=1\textwidth]{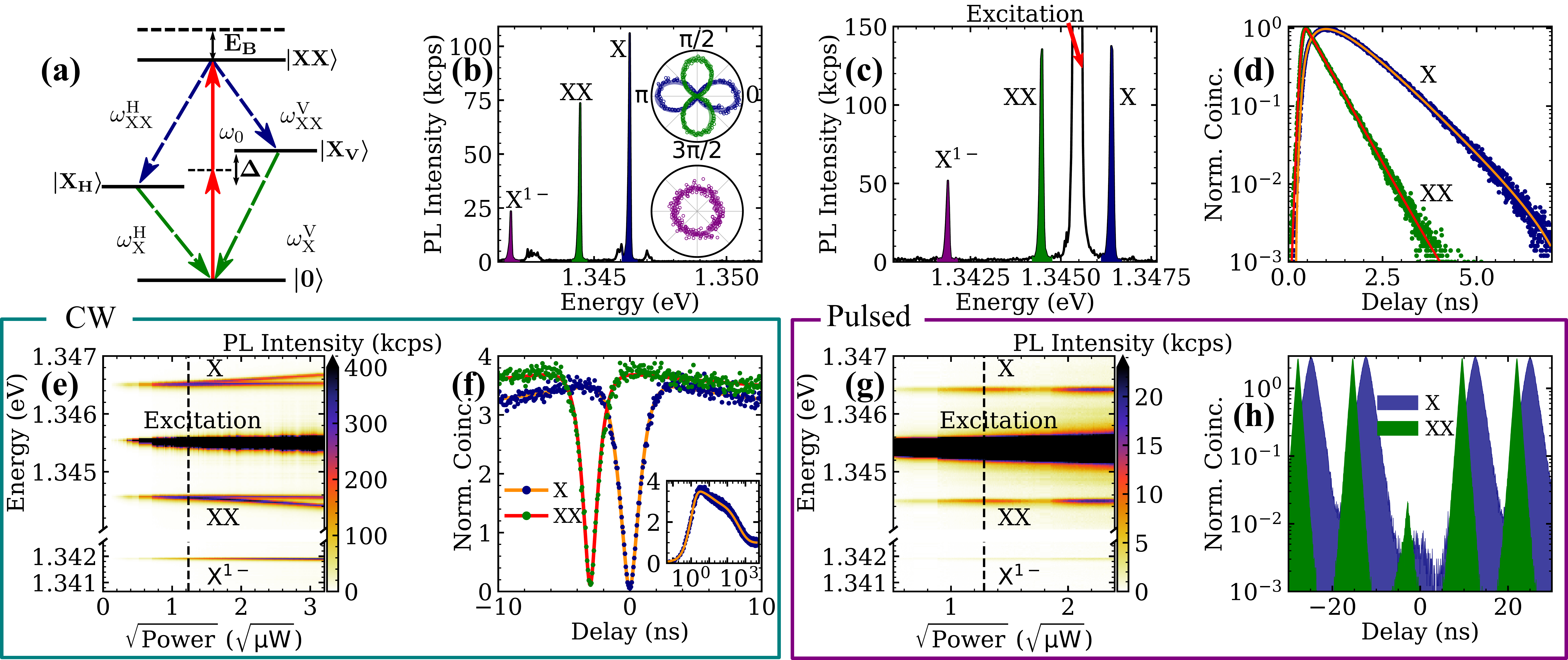}
    \caption{
        \textbf{Biexciton-exciton cascaded emission from a nanowire QD.}
        \textbf{(a)} Energy level schematic of the biexciton-exciton (XX-X) cascaded system. 
        $\Delta$: exciton fine structure splitting. $E_B$: biexciton binding energy. $\omega_0$: excitation energy under resonant two photon excitation. $\omega_\mathrm{X,XX}$: emission energy for exciton X and biexciton XX.
        \textbf{(b)} Emission spectra under above-band excitation (at 1.4938~eV), showing the emission lines from the X, XX and $\rm X^{1-}$ transition.
        Inset shows the emission energy as a function of polarization angle, confirming the nature of each emission line.
        \textbf{(c)} Emission spectra under resonant TPE (at $\omega_0=1.3455\,\mathrm{eV}$), reveals cleaner emission lines of X, XX and $\rm X^{1-}$. 
        \textbf{(d)} Lifetime measurement of the biexciton-exciton cascade under resonant two-photon excitation. 
        \textbf{(e)} Power dependence of the emission under CW TPE reveals an Autler-Townes splitting for both X and XX emissions.
        \textbf{(f)} Second-order intensity correlation histogram, $g^{(2)}$ of the X and XX emissions (at a power $1.49\,\mu\mathrm{W}$ indicated by the dashed line in (e)).
        Inset shows the same $g^{(2)}$ of the X emission, up to $5\,\mu \mathrm{s}$.  
        \textbf{(g)} Power dependence of the X and XX emissions under pulsed TPE reveals Rabi oscillations, with the $\pi$-pulse given by the black dashed line at $1.6\,\mu\mathrm{W}$. 
        \textbf{(h)} $g^{(2)}$ data for X and XX emissions under $\pi$-pulse excitation (with power indicated by the dashed line in (g)) in log scale.
        The $g^{(2)}$ data, normalized by the raw coincidence at $5\,\mu\mathrm{s}$, for XX emission in (d) and (f), are artificially shifted by -3~ns for clarity. 
    }
    \label{fig2}
\end{figure*}

Our candidate of choice to demonstrate spatially multiplexed streams of single photons from solid-state emitters is a sample of InAsP QDs embedded in deterministically positioned InP nanowires~\cite{dalacu_nanowire-based_2019}. 
Nanowire QDs offer an excellent coherent light-matter interface~\cite{cogan_depolarization_2018} and can emit high quality entangled photons~\cite{versteegh_observation_2014,jons_bright_2017}, while providing near unity coupling into optical fiber due to their wave-guiding properties~\cite{dalacu_ultraclean_2012,reimer_bright_2012} and Gaussian mode profile emission~\cite{bulgarini_nanowire_2014}. 
They are deterministically grown with up to 5 QDs on the same nanowire~\cite{laferriere_multiplexed_2020} and can be controllably positioned on integrated photonic chips ~\cite{zadeh_deterministic_2016,elshaari_strain-tunable_2018}.

To characterize our sample, we first perform coherent control of a single nanowire QD.
For this characterization, we replace the multi-spot microscope setup with a dark-field confocal microscope, consisting of a linear polarizer at the excitation and a quarter-wave plate and linear polarizer at the collection~\cite{kuhlmann_a_2013}. 
The linear polarizers are cross-polarized to suppress the excitation laser.
The scattered photons are then coupled into a single mode fibre. 
A free-space grating-based filtering setup with a bandwidth of $ \sim 100\,\mu\mathrm{eV}$ is used to spectrally filter the desired QD signal before HBT measurements. 
We record the time tags from each detector and extract the coincidences at various detection time delay, $\tau$~\cite{Ballesteros_readPTU_2019}. 

We perform spectroscopy on a single nanowire QD, focusing on the emission from the biexciton-exciton cascade.
The energy level schematic of the four-level cascade system is described in Figure~\ref{fig2}(a).
Upon excitation into the biexciton state, a cascaded radiative decay from biexciton $\ket{\mathrm{XX}}$ to vacuum ground state $\ket{0}$ is triggered via either of the two exciton states $\ket{\mathrm{X_{H,V}}}$.
This generates a pair of polarization-entangled photons, orthogonally polarized in the horizontal (H) and vertical (V) linear polarized basis. 

Focusing the CW above-band (1.4938~eV) excitation laser on a single nanowire, we obtain the photoluminescence (PL) emission spectrum from a single QD, shown in Figure~\ref{fig2}(b). 
The PL spectrum shows multiple emission lines, with the neutral exciton X, biexciton XX, and the negatively-charged exciton $\rm X^{1-}$ each exhibiting spectrometer-resolution-limited linewidths ($\mathrm{FWHM}=38\,\mu\mathrm{eV}$). 
We identify the nature of the emission lines via polarization resolved measurements. This is done by replacing the quarter-wave plate with a half-wave plate on the collection path and then, by rotating the half-wave plate with a piezoelectric rotator, recording the PL spectra at each rotation angle, $\theta$. We fit the data with $\omega_\mathrm{X, XX}=\omega_0+(\Delta/2)\sin(\theta+\phi)$, where $\omega_0$ is the mean emission energy, $\Delta/2$ is the amplitude and $\phi$ is the phase of the oscillation.
The amplitude extracted from the fit gives an exciton fine-structure splitting of $\Delta=11.2\,\mu \mathrm{eV}$, while the difference in the mean emission energy of the X and XX gives the biexciton binding energy $E_B=1.861\,\mathrm{meV}$. Here, we identify the XX and X pair via the out-of-phase oscillations in the emission energy as a function of linear polarization angle $\theta$, which implies linear and orthogonally polarized emissions.
The emission line corresponding to the $\rm X^{1-}$ does not have a fine structure splitting, and hence is circularly polarized. These results are shown as inset in Figure~\ref{fig2}(b).

Next, we demonstrate resonant two-photon excitation (TPE) on the same nanowire QD.
To perform TPE, we resonantly pump the population from the ground state $\ket{0}$ to the biexciton state $\ket{\mathrm{XX}}$ with an excitation energy set to exactly half of the energy difference between the two states.
This gives the TPE excitation energy of $\omega_0=E_\mathrm{XX-0}/2=\omega_\mathrm{X}-E_B/2=1.3455\,\mathrm{eV}$.
We then resolve one of the exciton fine structures $\ket{\mathrm{X_{H,V}}}$ by adjusting the linear polarizer in the collection to the polarization axis of the desired transition.
At the same time, the linear polarizer in the excitation is aligned orthogonal to the collection linear polarizer to suppress the excitation power.
The PL spectrum of the same QD under TPE is shown in Figure~\ref{fig2}(c).
The scattered excitation laser, indicated on the figure, is a result of imperfect polarization suppression of the laser in the dark-field microscope, mostly caused by the geometry of the tip of the nanowire.
Due to the resonant nature of the excitation, only the emission from the relevant transitions (X, XX and $\rm X^{1-}$) is observed.
In contrast with the PL spectrum in Figure~\ref{fig2}(b), the emission spectrum under TPE is much cleaner, indicating minimal excitation into the higher energy states.

Using the free-space spectral filter, we selectively filter the emission from either the X or XX transition.
Time-resolved measurements on both transitions, when exciting with a pulsed source ($\sim 3\,\mathrm{ps}$ pulse width, 80.3 MHz repetition rate, pulse area of $\pi$), give a radiative lifetime of $T_{1}^X=902\,(1)\,\mathrm{ps}$ and $T_1^{XX}=507\,(1)\,\mathrm{ps}$ for the X and XX emissions, respectively.
The results are shown in Figure~\ref{fig2}(d).
The fitting details for the time-resolved lifetime measurement are included in Appendix~\ref{sec:lifetime}.

Figure~\ref{fig2}(e) shows the power dependence in the PL emission spectra under continuous wave (CW) excitation.
At low excitation power, the intensity of the X and XX emissions increase with excitation power, until they reach saturation at $ 1.5\,\mu\mathrm{W}$, given by the black dashed line. 
Beyond the saturation, further increase in excitation power reveals Autler-Townes splittings in the emission of both X and XX transitions~\cite{jundt_observation_2008,gerardot_dressed_2009}.
This has been previously demonstrated in Refs.~\cite{ardelt_optical_2016,hargart_probing_2016,bounouar_path_2017} as a consequence of optical dressing of biexciton $\ket{\mathrm{XX}}$ and ground $\ket{0}$ states.
The formation of dressed states confirms the coherent nature of the excitation scheme, allowing for coherent manipulation of the emission from the biexciton-exciton cascade.

By fixing the excitation power at the saturation power, indicated by black dashed line in Figure~\ref{fig2}(e), we perform HBT measurement on the X and XX emissions.
Their normalized coincidence histogram $g^{(2)}$ are shown in Figure~\ref{fig2}(f).
The inset of Figure~\ref{fig2}(d) shows the long time scale, up to $5\,\mu\mathrm{s}$, of the normalized $g^{(2)}$ coincidences. 
The raw coincidences are normalized to the value at long time limit at $5\,\mu\mathrm{s}$, where it is unaffected by the bunching in the $g^{(2)}$.
Accounting for the multiple bunching time scales present in the $g^{(2)}$ measurement, we fit the data with
\begin{equation}
g^{(2)}_\mathrm{X\,(XX)}(\tau) = g^{(2)}_\mathrm{RF}(\tau)\prod_{j\in\{a,b,c\} } \left(1+j\,e^{-\tau/\tau_j}\right)
\label{eqn:g2}
\end{equation}
where $a$, $b$ and $c$ are the bunching amplitudes for the corresponding bunching time scales $\tau_a$, $\tau_b$ and $\tau_c$.
The theoretical function $g^{(2)}_\mathrm{RF}(\tau)$ follows the expression for a resonantly driven two-level system~\cite{scully_quantum_1997}.
The resulting $g^{(2)}$ fit function is the convolution of Eq.~\ref{eqn:g2} with a Gaussian instrument response function ($\mathrm{FWHM}=160\,\mathrm{ps}$).
The fit shows excellent agreement with the experimental data, giving suppressed multi-photon emission probability of $g_\mathrm{X}^{(2)}(0) = 0.006\,(1)$ and $g_\mathrm{XX}^{(2)}(0) = 0.023\,(2)$. 
The fit also reveals different bunching amplitudes and time scales for both X and XX emissions.
The bunching in $g^{(2)}$ is related to spectral fluctuations of the emitter, likely due to charge noise in the QD environment and at the nanowire surface. The fit values of the bunching amplitudes and time scales of the $g^{(2)}$ for both X and XX emissions are summarized in Table~\ref{tab:table1}.

\begin{table}[ht]
    \caption{\label{tab:table1}%
    The bunching time scale and amplitude, extracted from the fit to the $g^{(2)}$ data for both exciton X and biexciton XX emissions.
    The error, given by the standard deviation extracted from the fit, is included in the bracket. 
    }
    \begin{ruledtabular}
    \begin{tabular}{ l c c}
    \textrm{}&
    \textrm{Exciton, X} &
    \textrm{Biexciton, XX}\\
    \colrule
    $\tau_a\,\mathrm{(ns)}$ & 12 (1) & 19 (1)\\
    $\tau_b\,\mathrm{(ns)}$ & 391 (3) & 376 (1)\\
    $\tau_c\,\mathrm{(ns)}$ & 769 (6) & 992 (9)\\
    $a$ & 0.193 (4) & 0.100 (2)\\
    $b$ & 0.717 (16) & 1.432 (12)\\
    $c$ & 0.789 (17) & 0.408 (8)\\
    \end{tabular}
    \end{ruledtabular}
\end{table}

Figure~\ref{fig2}(g) shows the power dependence in the PL emission spectra under pulsed resonant two-photon excitation. By varying the average excitation power of the pulsed source, we observe clear Rabi oscillations in the intensity of both X and XX emissions.
We fit the data with a damped sinusoidal function to extract the power that corresponds to $\pi$ pulse at $P_\pi = 1.6\,\mu \mathrm{W}$.
We attribute the increased in counts in the $3\pi$ pulse as a result of experimental imperfection, most likely originated from pulse chirping~\cite{glassl_biexciton_2013,kaldewey_demonstrating_2017} or slight detuning from the ideal resonant two-photon excitation energy, $\omega_0$~\cite{ardelt_dissipative_2014,bounouar_phonon-assisted_2015}.

Next, we perform a pulsed HBT measurement by exciting the QD using a $\pi$-pulse ($P_{\pi}=1.6\,\mu\mathrm{W}$).
The result is shown in Fig.~\ref{fig2}(h) where we obtain similar multi-photon emission probability of $g^{(2)}(0) = 0.0117\,(4)$ and $g^{(2)}(0) = 0.0291\,(5)$ for X and XX emission, respectively.
The raw coincidences are normalized to the peaks at $ 5 \, \mu \mathrm{s}$, giving a long bunching time scale of $\approx 300\,\mathrm{ns}$.
Nevertheless, the observation of Rabi oscillation and vanishing $g^{(2)}(0)$ demonstrate coherent control of the state population as well as highly pure single photon emission from the nanowire QDs. 

\section{Multiplexed single photon sources}
\begin{figure*}
    \includegraphics[width=1\textwidth]{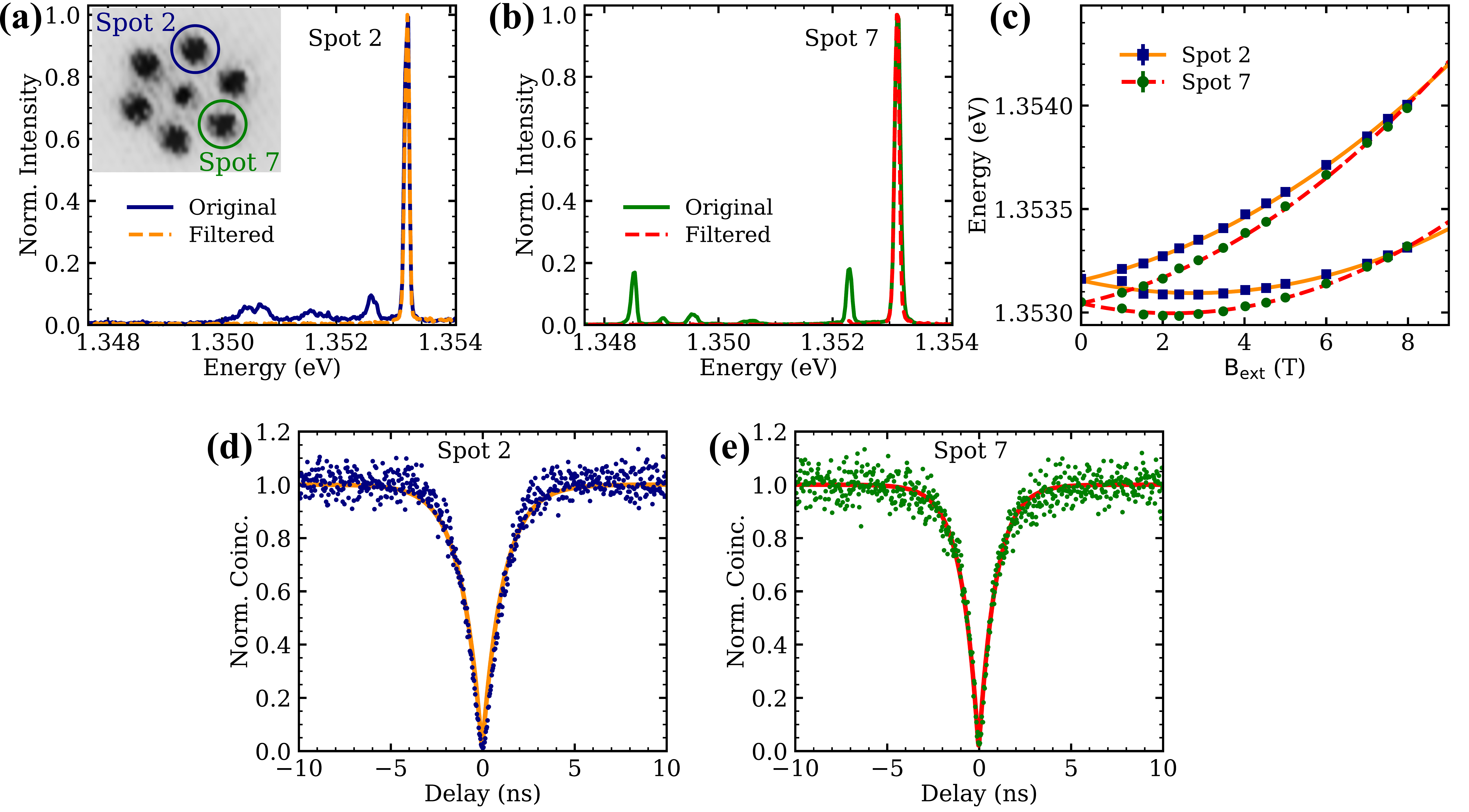}
    \caption{
        \textbf{Multiplexing single photons from independent quantum emitters.}
        \textbf{(a, b)} Emission spectra of the original (solid line) and filtered (dashed line) from wires at Spot 2 (c) and Spot 7 (d), under above-band excitation.
        The inset figure indicates the excitation and collection spots for wires at Spot 2 and Spot 7.
        \textbf{(c)} Emission energy of the filtered QDs as a function of external magnetic field, $\mathrm{B_{ext}}$. 
        The fits give a g-factor and diamagnetic shift coefficient of $1.53\,(1)$, $7.99\,(7)\,\mu\mathrm{eV/T^2}$ for the QDs at Spot 2 (solid line) and $1.49\,(1)$, $9.65\,(8)\,\mu\mathrm{eV/T^2}$ at Spot 7 (dashed line).
        \textbf{(d, e)} Normalized $g^{(2)}$ for the emission from nanowires at Spot 2 (f) and Spot 7 (g), at $\mathrm{B_{ext}}=0$, showing a suppressed multi-photon emission of $g^{(2)}(0)\sim 0$.
    }
    \label{fig3}
\end{figure*}

Having demonstrated that nanowire QDs can generate high-purity single photons, we now aim to illustrate the full functionality of our microscope by multiplexing the single photon streams of two degenerate QDs. 
Here, the ability to probe multiple nanowires in parallel accelerates the search for pairs of QDs emitting at similar emission energy in the nanowire arrays. We address the nanowires in the multi-spot microscope using above-band (1.4938~eV) excitation, and find two QDs with similar exciton emission (centered at 1.3531~eV) with an energy difference of 0.113~meV. The emission spectra for both of the QDs, before (original) and after (filtered) spectral filtering, are shown in Figure~\ref{fig3}(a) and (b). 
These QDs are labeled by their excitation spot (Spot 2 and Spot 7), as indicated in the inset of Figure~\ref{fig3}(a). By applying an external magnetic field, $\mathrm{B_{ext}}$, we can fine tune and minimize the difference in the emission energies of both QDs. This result is shown in Figure~\ref{fig3}(c), where their emission energies match at $\mathrm{B_{ext}}\approx 8\,\mathrm{T}$.
The fits yield a g-factor of 1.53 (1.49) and diamagnetic shift coefficient of $7.99\,(9.65)\,\mu \mathrm{eV/T^2}$ for the QD at Spot 2 (Spot 7).

To verify high-purity multiplexed single photon emission from these QDs, we perform HBT measurements on the emissions from the two QDs. 
Their corresponding coincidence histogram, $g^{(2)}$ are shown in Figure~\ref{fig3}(d, e). 
Fitting the $g^{(2)}$ data with Eq.~\ref{eqn:g22}
\begin{equation}
    g^{(2)}(\tau) = 1- \xi \exp(-|\tau|/T_1),
\label{eqn:g22}
\end{equation}
gives vanishing multi-photon emission probability of $g^{(2)}(0)=1-\xi\sim 0 $ for both QDs at Spot 2 and Spot 7.
The decay times, $T_1$ extracted from the $g^{(2)}$ histograms are similar for both QDs: $T_1=1.14\,(1)\,\mathrm{ns}$ for QD at Spot 2 and $T_1=0.92\,(1)\,\mathrm{ns}$ for QD at Spot 7.
These results certify the functionality of this microscope, as we demonstrate high purity single photon multiplexing from emitters on the same chip.

\section{Discussion and outlook}
A few straightforward improvements could be made to our multi-spot microscope setup. One issue to address is the strong chromatic aberrations introduced by the collimating and relay lenses, which causes the light from the above-band excitation laser to be defocused at the plane of the nanowires (optimized for the collection wavelength) and leads to cross-talk between excitation spots. To improve this, a cryostat that allows a large field of view without relay lenses (e.g. the Montana Cryostation or Attocube attoDry800) or achromatic lenses could be used. A second issue: the multi-mode behavior of the commercial MCFs at the QD emission wavelengths (910 -- 950~nm) leads to extra cross-talk, poor excitation laser extinction in the dark-field configuration, and losses when coupling the MCF into single mode fibers for interferometry or detection. Custom fabricated MCFs which are single mode at the QD wavelength would alleviate these problems. Finally, an increase in the number of cores in the MCF is desirable to maximize scalability. 

Nevertheless, we have successfully demonstrated multiplexing of single photons from degenerate nanowire QDs using the novel multi-spot microscope. The nanowire QDs exhibit near perfect suppression of multi-photon emission under above-band and resonant two-photon excitation, albeit the observation of blinking emission with a time scale up to $\sim 1\,\mu\mathrm{s}$ due to presence of spectral fluctuations under resonant excitation.
These spectral fluctuations, likely originating from charge noise from the solid-state environment and at the surface of the nanowire, degrade the performance of the photon source and prevent high visibility two-photon interference experiments.

Hence, multiplexing indistinguishable single or entangled photon streams would benefit from optimized quantum emitter devices, which would ideally include deterministic positioning, near perfect extraction efficiency, in-situ tunability of each emitter's emission energy, minimal spectral fluctuations, and Purcell-enhanced emitter lifetimes. Additionally, easy incorporation of coherent optical control is necessary. Our choice of nanowire QDs, motivated here by the deterministic positioning and high extraction efficiency, can potentially be individually tuned using strain ~\cite{kremer_strain-tunable_2014,chen_controlling_2016,fiset-cyr_-situ_2018}, but the dangling bonds at the nanowire sidewalls likely will remain an obstacle to reduce charge noise and the nanowire geometry hinders resonant laser excitation in the dark-field configuration. 
Ideally, the emitters would be incorporated into a heterostructure device with individual contacts for each spatial position in the array to provide charge control, control of the emitter energy, charge environment stabilization, and suppression of solid-state environmental noise~\cite{kuhlmann_charge_2013,kuhlmann_transform-limited_2015}. While micro-pillar and bullseye type devices based on self-assembled QDs ~\cite{ding_-demand_2016,somaschi_near-optimal_2016,liu_solid-state_2019} offer high-extraction efficiency and strong Purcell enhancement, deterministic positioning remains a significant challenge to overcome. On the other hand, waveguide architectures which can can couple light into and out of remotely located quantum emitters could be ideal. Such a platform offers high extraction efficiency, significant Purcell enhancement, charge control, the ability for local strain tuning, and the freedom to guide light anywhere on chip to and from randomly positioned QDs~\cite{grim_scalable_2019,uppu_scalable_2020}. However, generation of high visibility two-photon interference from remote QDs is a significant challenge to be met in all approaches~\cite{patel_two_2010,reindl_phonon-assisted_2017,kambs_limitations_2018}. Regardless of the chosen nanophotonics platform, our novel multi-spot microscope enables genuine scalability by providing individual excitation and collection to multiplex streams of indistinguishable single photons for optical quantum information processes.

\begin{acknowledgments}
This work was supported by the EPSRC (Grants No. EP/L015110/1, EP/M013472/1, and EP/P029892/1), the ERC (Grant No. 725920), and the EU Horizon 2020 research and innovation program under grant agreement no. 820423. 
B.D.G. thanks the Royal Society for a Wolfson Merit Award and the Royal Academy of Engineering for a Chair in Emerging Technology.
\end{acknowledgments}

\bibliography{reference}

\newpage
\appendix
\section{Lifetime Measurements}\label{sec:lifetime}
We proceed to fit the coincidence histogram from the time-resolved lifetime measurements with the following function:
\begin{align}
\label{sup2}
\mathcal{I}_1(\tau,\Gamma_1,\sigma) &= \exp\left( \frac{\Gamma_1(\Gamma_1\sigma^2-2\tau)}{2}\right)\mathrm{erfc}\left(\frac{\Gamma_1\sigma^2-\tau}{\sqrt{2}\sigma}\right),\\
\mathcal{I}_2(\tau,\Gamma_1,\Gamma_2,\sigma) &= A\,\mathcal{I}_1(\tau,\Gamma_1,\sigma) + B\,\mathcal{I}_1(\tau,\Gamma_2,\sigma).
\label{sup3}
\end{align} 
Here, $\mathcal{I}_1(\tau,\Gamma_1,\sigma)$ is simply a Heaviside step function, defined such that at positive time, the function is consists of a single exponential functions, with decay time $\tau_1=1/\Gamma_1$, convoluted with the instrument response function (modelled as a Gaussian with standard deviation $\sigma$). 
Fitting parameters $A$ and $B$ in $\mathcal{I}_2(\tau,\Gamma_1,\Gamma2,\sigma)$ represent the weighted contribution of the two exponential decays.

For exciton, X emission, the data is fitted with Eq.~\ref{sup3} which consists of two exponential decays, with time scales $T_\mathrm{X}$ and $T_\mathrm{XX}$. 
For biexciton, XX emission, the data is fitted with the same equation but with $B=0$ as the only relevant time scale is the biexciton radiative lifetime, $T_\mathrm{XX}$.

\end{document}